\documentclass[reprint,aip,graphicx]{revtex4-1}

\pdfoutput = 1

\usepackage{amsmath}
\usepackage{amssymb}

\draft 

\usepackage{graphicx}

 \usepackage{multirow} 

\begin{document}


\title{Narrow-bandwidth solar upconversion: design principles, efficiency limits, and case studies}



\author{Justin A. Briggs}
\email[Corresponding author: ]{jabriggs@stanford.edu}
\affiliation{Department of Applied Physics, Stanford University}
\affiliation{Department of Materials Science and Engineering, Stanford University}
\author{ Ashwin C. Atre}
\affiliation{Department of Materials Science and Engineering, Stanford University}
\author{Jennifer A. Dionne}
\affiliation{Department of Materials Science and Engineering, Stanford University}


\date{\today}

\begin{abstract}
We employ a detailed balance approach to model a
single-junction solar cell with a realistic narrow-band,
non-unity-quantum-yield upconverter. As upconverter bandwidths are
increased from 0 to 0.5 eV, maximum cell efficiencies increase from
the Shockley-Queisser limit of 30.58\% to over 43\%. Such
efficiency enhancements are calculated for upconverters with
near-infrared spectral absorption bands, readily accessible with
existing upconverters. While our model shows that current
bimolecular and lanthanide-based upconverting materials will
improve cell efficiencies by $<$1\%, cell efficiencies can
increase by several absolute percent with increased upconverter
quantum yield - even without an increased absorption bandwidth. By
examining the efficiency limits of a highly realistic solar
cell-upconverter system, our model provides a platform for
optimizing future solar upconverter designs.\end{abstract}

\pacs{}

\maketitle 

	In one hour the sun delivers enough energy to the earth to meet global needs for an entire year \cite{LewisAndNocera}. However, semiconductor-based solar technologies are generally unable to utilize photons below the device bandgap and can thus harvest only a small portion of this energy. Spectral upconverters provide a possible solution. Placed behind a cell, they capture transmitted sub-bandgap photons and convert them to a frequency range that can be utilized by the cell. Because they are electrically isolated from the solar cell, they neither introduce recombination pathways for electron-hole pairs (as intermediate level systems do \cite{KeeversAndGreen,LuqueAndMarti,GuttlerAndQueisser}), nor require current matching (as multi-junction systems do \cite{MultiJunction1,MultiJunction2,MultiJunction3}). Upconversion has been observed in many systems, including rare earth and transition metal ions \cite{RareEarth1,RareEarth2,RareEarth3,RareEarth4}, various semiconductor quantum dots\cite{QD1,QD2,QD3}, and metallated macrocycles \cite{Macro1,Macro2,Macro3,Macro4,Macro5}. 
Further, upconversion quantum efficiencies greater than 20\% have been reported for solid state systems under low-power, non-coherent illumination \cite{Castellano20percent}. 

Trupke et al. considered the theoretical solar cell efficiency improvements possible with the addition of an ideal upconverter \cite{Trupke2002}, and Atre and Dionne extended this analysis to account for cell non-idealities and a non-radiative relaxation pathway in the upconverter \cite{Ashwin}. In a recent study, Johnson and Conibeer examined the impact of an idealized upconverter on a highly realistic c-Si solar cell \cite{Johnson_cSi}. While these works assumed that the upconverter was able to absorb over the entire sub-bandgap spectrum, all known upconverters are found to absorb and emit radiation in relatively narrow bands, ranging from less than .1 eV to over .4 eV \cite{CastellanoReview, Castellano20percent, IR806andLanth}. It is necessary to consider this practical limitation in order to accurately predict the efficiency enhancements afforded by upconversion.

    Here, we theoretically study a solar cell with
an upconverter characterized by narrow-band absorption and
emission. We investigate the expected efficiency limits of a highly
realistic solar cell-upconverter system, both optimizing
upconverter design for a given cell bandgap and exploring the
efficiencies possible with existing and next-generation upconversion
systems.

	We consider an ideal single-junction solar cell with an electrically isolated upconverter behind it, as depicted in Fig. 1(a). The upconverter is modeled as two low bandgap solar cells ($C_3$ and $C_4$) in series with a high bandgap photodiode ($C_2$)  \cite{Trupke2002}. As seen in Fig. 1(b), electrons excited by low energy photons in the low bandgap cells drive the photodiode into forward bias, allowing it to radiate above-bandgap photons. A generalized Planck radiation law \cite{WurfelPlanck} is used to describe the photon flux density emanating from the solar cell, the upconverter, the sun, and the ambient environment: 

\begin{equation}\label{FluxDensity} 
j(T,\mu,\epsilon,\hbar\omega) = \frac{\epsilon}{4 \pi^3 \hbar^3 c^2} \frac{(\hbar\omega)^2}{exp\left(\frac{\hbar\omega - \mu}{kT}\right) - 1}
\end{equation}

\noindent
Here, $\hbar$ is the reduced Planck's constant, $c$ is the speed of light in vacuum, $T$ is the temperature of the radiating body, $\mu$ is the operating voltage of the body times the elementary charge, and $\epsilon$ is the \'{e}tendue factor accounting for geometrical concentration of light and reflection from interfaces with mismatched refractive indices \cite{Badescu2007}. 

To enforce the narrow-band absorption and emission characterizing known upconverters, we impose a spectral weighing lorentzian function $f(A,W,\hbar\omega)$, parameterized by a centroid $A$ and a full width at half max $W$. The photon fluxes from each cell comprising the upconverter are computed as the convolution of the Planck distribution and the appropriate weighting function: 

\begin{align} \label{WeightedFlux}
& \dot{N_i}(E_l,E_u,T_i,\mu_i,A_i,W_i,\epsilon) = \nonumber  \\
& \int_{E_l}^{E_u} j(T_i,\mu_i,\epsilon,\hbar\omega) f(A_i,W_i,\hbar\omega) d(\hbar\omega)
     \end{align}

\noindent
where the subscript $i$ refers to the $i^{th}$ cell. Photon fluxes from the sun, the environment, and the solar cell are given directly by the integral of the spectral density:

\begin{align} \label{Flux}
& \dot{N_i}(E_l,E_u,T_i,\mu_i,\epsilon) = \nonumber \\
& \int_{E_l}^{E_u} j(T_i,\mu_i,\epsilon,\hbar\omega) d(\hbar\omega)
\end{align}

\begin{figure}
\includegraphics[width=.5\textwidth]{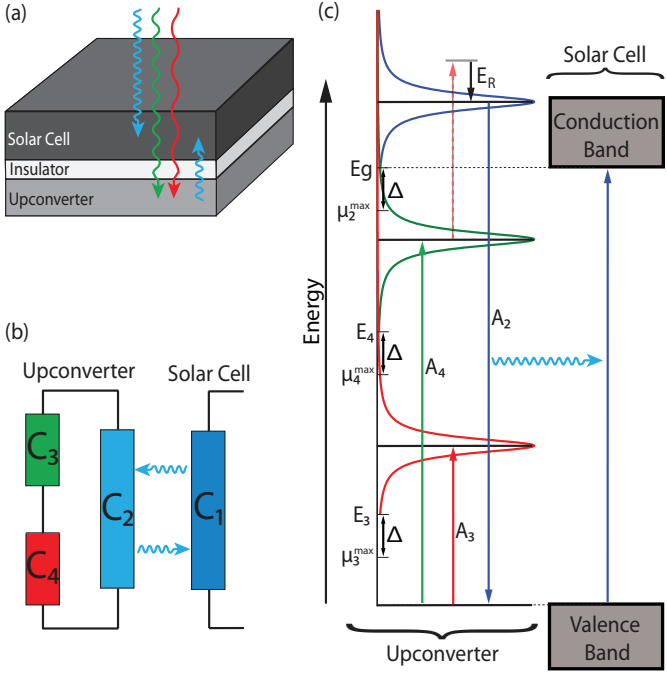}%
\caption{Schematics outlining the solar cell/upconverter system under consideration. (a) Above-bandgap light is absorbed by the solar cell, which is electrically isolated from the upconverter. Sub-bandgap light is transmitted by the solar cell, absorbed by the upconverter, and re-radiated back into the cell as above-bandgap light. (b) The upconverter is equivalent to a circuit containing two low bandgap solar cells (C$_3$ and C$_4$) in series with a high bandgap photodiode (C$_2$). (c) Energy level diagram indicating the relevant transitions. The maximum operating potentials for cells C$_2$, C$_3$, and C$_4$ are given by $\mu_2^{max}$, $\mu_3^{max}$, and $\mu_4^{max}$; $E_g$, $E_3$, and $E_4$ represent the effective band edges for cells C$_2$, C$_3$, and C$_4$, respectively; $\Delta = \frac{3}{2}kT$ is a thermalization energy giving the average relationship between the maximum operating potentials and the effective band edges; $A_2$, $A_3$, and $A_4$  are the centroids of the upconverter emission and absorption lorentzians. The solar cell conduction and valence bands extend to positive and negative infinity, respectively.}%
\end{figure}

 	An energy level diagram representing the resulting three-level system is shown in Fig. 1(c). $A_2$, $A_3$, and $A_4$ are the centroids of the emission and absorption lorentzians associated with cells $C_2$, $C_3$, and $C_4$, while $\mu_2^{max}$, $\mu_3^{max}$, and $\mu_4^{max}$ are their maximum operating potentials. The parameters $E_3$ and $E_4$ represent the effective band edges associated with the low energy upconverter transitions, and thus define the lower bounds of integration when Eq. (2) is applied to the low bandgap cells, $C_3$ and $C_4$. Minimum energy inter-band transitions in a cell with maximum operating potential $\mu_i^{max}$ occur, on average, for electrons lying $\frac{3}{2}kT$ above the conduction band edge, so $(\mu_i^{max} + \frac{3}{2}kT) = E_i$ is the approximate lower bound for the energy of photons emitted from such a cell. Similarly, $(\mu_2^{max} + \frac{3}{2}kT) = E_g$ is the effective band edge associated with cell $C_2$. 
	
	Conservation of energy dictates that $A_3+A_4-E_{R} = A_2$, where $E_R$ is a non-radiative relaxation energy. We assume that this relaxation energy accounts for all non-radiative loss in the system and therefore allow a single lorentzian to characterize both the absorption and emission of each cell comprising the upconverter. Further, we assume that the bandwidths of all absorption and emission processes in the upconverter are equal, and refer to this value as the upconverter bandwidth. To ensure good optical coupling between the upconverter and the solar cell we select an upconverter emission energy $A_2$ that is sufficiently above ($>$.1 eV) the cell bandgap.

	We employ a generalization of the detailed balance approach used by Shockley and Queisser \cite{ShockleyQueisser,TrupkeGreen,Badescu2008}. In this model the current density in the $i^{th}$ cell is equal to the elementary charge times the difference between the photon fluxes absorbed and emitted by that cell:

\begin{equation}\label{Current} 
I_i = q(f_{abs,i}\dot{N}_{inc,i} - f_{rec,i}\dot{N}_{em,i})
\end{equation}

\noindent
Here, $f_{abs,i}$ and $f_{rec,i}$ are the absorption efficiency and the inverse of the radiative recombination efficiency of the $i^{th}$ cell \cite{Ashwin}. The currents in the solar cell ($C_1$) and each cell representing the upconverter ($C_2,C_3,C_4$) are computed as:

\begin{align} \label{I1}
& I_1/q = \nonumber
\\ &f_{abs,C_1} \nonumber
\\ & \quad \times [ \dot{N}(E_g,\infty,T_S,0,\epsilon_{S \to SC}) \nonumber
\\  & \quad +\dot{N}(E_g,\infty,T_A,0,\epsilon_{A \to SC}) \nonumber
\\  & \quad +\dot{N}(E_g,\infty,T_A,\mu_2,A_2,W_2,\epsilon_{UC \to SC}) ]
\\  &-f_{rec,C_1}[ \dot{N}(E_g,\infty,T_A,\mu_1,\epsilon_{SC \to A}) \nonumber
\\  & \quad +\dot{N}(E_g,\infty,T_A,\mu_1,A_2,W_2,\epsilon_{SC \to UC})] \nonumber
\end{align}

\begin{align} \label{I2}
&I_2/q =  \nonumber
\\    &         f_{abs,C_2}\{ \dot{N}(E_g,\infty,T_A,\mu_1,A_2,W_2,\epsilon_{SC \to UC}) \nonumber
\\    &    \quad  +[1-f_{abs,C_1}] \nonumber
\\    & \quad \quad \times [\dot{N}(E_g,\infty,T_S,0,A_2,W_2,\epsilon_{S \to SC \to UC}) \nonumber
\\    &  \quad \quad +\dot{N}(E_g,\infty,T_A,0,A_2,W_2,\epsilon_{A \to SC \to UC}) ] \}
\\    &        -f_{rec,C_2} \nonumber
\\    & \quad \times \{ f_{abs,C_1}\dot{N}(E_g,\infty,T_A,\mu_2,A_2,W_2,\epsilon_{UC \to SC}) \nonumber
\\    &    \quad  +[1-f_{abs,C_1}] \nonumber
\\    & \quad \quad \times \dot{N}(E_g,\infty,T_A,\mu_2,A_2,W_2,\epsilon_{UC \to SC \to A})  \} \nonumber
\end{align}

\begin{align} \label{I34}
& I_{3,4}/q  =   \nonumber
\\      &       f_{abs;C_3,C_4}\{ \dot{N}(E_{3,4},E_g,T_S,0,A_{3,4},W_{3,4},\epsilon_{S \to SC \to UC}) \nonumber
\\      &      +\dot{N}(E_{3,4},E_g,T_A,0,A_{3,4},W_{3,4},\epsilon_{A \to SC \to UC}) \nonumber
\\      &      +[1-f_{abs,C_1}] \nonumber
\\      & \quad \times [\dot{N}(E_g,\infty,T_S,0,A_{3,4},W_{3,4},\epsilon_{S \to SC \to UC}) \nonumber
\\      &   \quad   +\dot{N}(E_g,\infty,T_A,0,A_{3,4},W_{3,4},\epsilon_{A \to SC \to UC}) ] \}
\\\     &     -f_{rec;C_3,C_4} \nonumber
\\      & \quad \times \{ \dot{N}(E_{3,4},E_g,T_A,\mu_{3,4},A_{3,4},W_{3,4},\epsilon_{UC \to SC \to A}) \nonumber
\\      &     \quad +[1-f_{abs,C_1}]  \nonumber
\\      & \quad \quad \times \dot{N}(E_g,\infty,T_A,\mu_{3,4},A_{3,4},W_{3,4},\epsilon_{UC \to SC \to A})   \nonumber
\\      &   \quad  +f_{abs,C_1}\dot{N}(E_g,\infty,T_A,\mu_{3,4},A_{3,4},W_{3,4},\epsilon_{UC \to SC})  \} \nonumber
\end{align}
\\

\noindent
Here, $T_S$ and $T_A$ are the temperatures of the sun and the ambient environment (assumed to be 6000K and 300K, respectively), $\mu_i$ is the operating potential of the $i^{th}$ cell, and the \'{e}tendue factor $\epsilon_{X \to Y \to Z}$ accounts for reflection from interfaces as light passes from layer X to Y to Z (where $S$, $A$, $SC$, and $UC$ indicate the sun, the ambient environment, the solar cell, and the upconverter). We do not explicitly assume photon selectivity, but radiative exchange between the solar cell and cells $C_3$ and $C_4$ (e.g. radiation emitted from cell $C_3$ and absorbed by the solar cell) is suppressed by the rapid falloff of the Lorentzian weighting functions. As such, terms accounting for this coupling are typically multiple orders of magnitude smaller than other relevant terms, effectively enforcing photon selectivity. 

	Using the above thermodynamic model and detailed balance equations, we optimize the solar cell efficiency as a function of the upconverter absorption peak positions and the operating potentials of the solar cell and upconverter cells. This optimization is subject to the series circuit constraints placed on the currents and operating potentials of the cells comprising the upconverter:

\begin{equation} \label{currents}
I_2  = -I_3 = -I_4
\end{equation}

\begin{equation} \label{potentials}
\mu_2 = \mu_3 + \mu_4
\end{equation}
\\
\\
\indent
We begin by assuming both the solar cell and the upconverter have unity absorption and radiative recombination efficiencies. We calculate the maximum power conversion efficiency of this idealized system as a function of cell bandgap and upconverter bandwidth (Fig. 2(a)). In the limit of zero bandwidth, i.e. no upconversion, the Shockley-Queisser limit is recovered, giving 30.58\% cell efficiency at a bandgap of 1.3 eV. As the bandwidth increases, the peak efficiency increases and the cell bandgap at which this maximum is achieved blueshifts. The addition of ideal (i.e., unity quantum efficiency) .1 eV, .3 eV, and .5 eV bandwidth upconverters to an ideal 1.7 eV bandgap solar cell results in an efficiency increase from 28.24\% to 33.52\%, 39.94\%, and 43.58\%, respectively. As .5 eV is approximately an upper limit for the bandwidth of known upconverters, this later value represents the maximum expected efficiency enhancement possible with a narrow-bandwidth upconverter.
\\
\indent
This analysis employs an optimized relaxation energy, which
we compute to be .48 eV. Fig. 2(b) depicts the absolute increase
in cell efficiency over the Shockley-Queisser curve as a function
of upconverter relaxation energy for a .1 eV bandwidth upconverter. As seen, this value is largely
independent of cell bandgap. As the upconverter bandwidth is
increased (data not shown), the optimal relaxation energy becomes
smaller, but the increase in cell efficiency is not significantly
affected in the narrow-bandwidth regime.

\begin{figure}[h!]
\centering
\includegraphics[width=.5\textwidth]{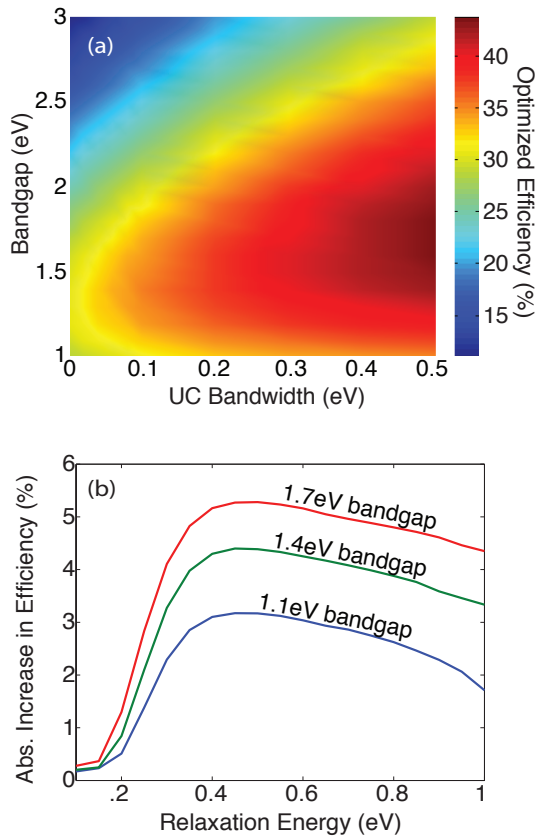}\\
\caption{Ideal solar cell with an ideal upconverter. (a) The maximum power conversion efficiency is calculated as a function of cell bandgap and upconverter bandwidth. The Shockley-Queisser limit is thus generalized to include a narrow-band upconverter. (b) The increase in cell efficiency is computed as a function of upconverter relaxation energy for a .1 eV bandwidth upconverter.}
\end{figure}

Our narrow-bandwidth model allows us to optimize the spectral characteristics of an upconverter. To fully explore this capability we employ the AM1.5G solar irradiance spectrum \cite{NREL}, which takes into account atmospheric absorption of solar radiation. Using this empirical spectrum in lieu of the solar Planck distribution, we optimize the solar cell efficiency as a function of the spectral locations of the upconverter absorption peaks. 

As can be seen in Fig. 3, the optimal upconverter absorption peaks in a general
dual-absorber system tend to increase with cell bandgap and are
generally in the near-infrared. For example, for a 1.7 eV bandgap,
characteristic of amorphous silicon and some organic photovoltaics,
the optimal absorption peaks lie at 1.02 eV and 1.47 eV. It is
encouraging that these values are readily accessible with existing
upconverters \cite{IR806andLanth, Polman_UC}. Note however that the ideal absorption positions depend
strongly on the absorption lines in the AM1.5G spectrum.


\begin{figure}[h!]
\centering
\includegraphics[width=.5\textwidth]{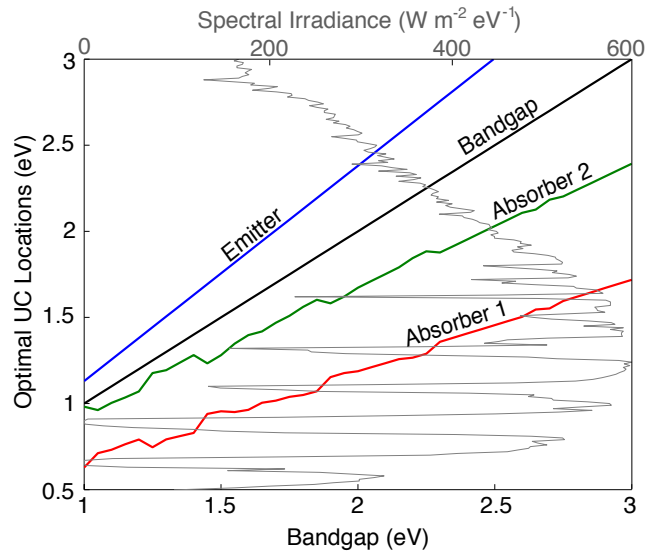}\\
\caption{The AM1.5G solar irradiance spectrum is used to calculate the optimal spectral locations for the low energy absorption peaks in the upconverter. The emitter position is fixed relative to each bandgap. The AM1.5G spectrum is plotted (relative to the upper horizontal axis) for reference.}
\end{figure}

As an application of our model, we explore the efficiency gains possible with the application of two existing upconverters. We consider both a bimolecular system (the absorber Pd(II) octaethylporphyrin  (PdOEP) paired with the emitter diphenylanthracene (DPA)) \cite{Castellano20percent}, and a lanthanide-based system (oleylamine-coated $\beta$-NaYF$_4$:Yb,Er nanoparticles sensitized with the carboxylated cyanine dye IR806)\cite{IR806andLanth}. The spectral characteristics of these systems are summarized in Table I. We chose these upconveters because they show particular promise: the bimolecular system has achieved upconversion efficiencies over 20\%\cite{Castellano20percent}, while the lanthanide-based system has energetics favorable for existing photovoltaic technologies \cite{IR806andLanth}.

\renewcommand{\thetable}{\Roman{table}}
\begin{table}[h!]
\begin{center}
\begin{tabular}{cc|c|c|c|c|l}
\cline{3-6}
& & \multicolumn{2}{c|}{bimolecular}  & \multicolumn{2}{c|}{lanthanide} \\ 
\cline{3-6}
     & & PdOEP & DPA & IR806 &  Er$^{3+}$ \\ \cline{1-6} 
\multicolumn{1}{|c}{\multirow{2}{*}{peak(eV)}} &
\multicolumn{1}{|c|}{abs.} & 2.28 & 3.30 & 1.54 & 1.28 &     \\ \cline{2-6}
\multicolumn{1}{|c}{}                        &
\multicolumn{1}{|c|}{em.} & 1.86 & 2.80 & 1.49 & 2.27 &     \\ \cline{1-6}
\multicolumn{1}{|c}{\multirow{2}{*}{FWHM(eV)}} &
\multicolumn{1}{|c|}{abs.} & 0.07 & 0.68 & 0.14 & 0.03 &  \\ \cline{2-6}
\multicolumn{1}{|c}{}                        &
\multicolumn{1}{|c|}{em.} & 0.10 & 0.39 & 0.15 & 0.12 &  \\ \cline{1-6}
\end{tabular}
\end{center}
\caption{Spectral parameters (absorption and emission peaks and full widths at half max [FWHMs]) characterizing the two upconverting systems under study, the bimolecular system (PdOEP+DPA) and the lanthanide-based system (IR806-sensitized oleylamine-coated $\beta$-NaYF$_4$:Yb,Er nanoparticles).}
\end{table}

\begin{figure}[h!]
\centering
\includegraphics[width=.5\textwidth]{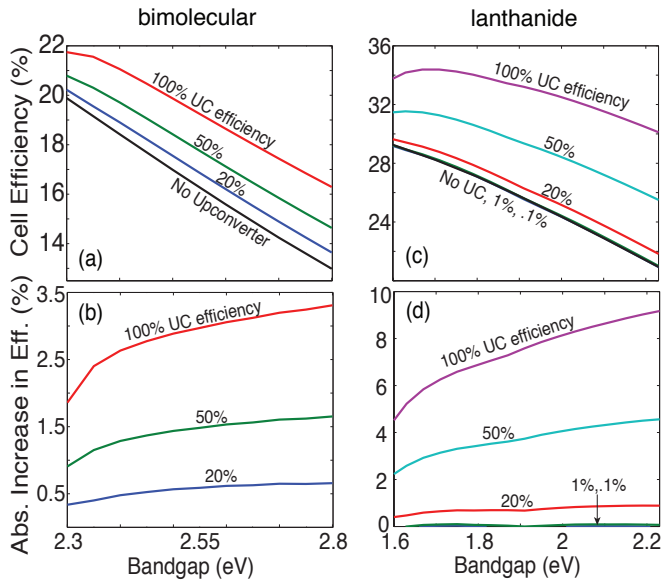}\\
\caption{Case studies of existing upconverting systems with different upconverter efficiencies shown. The raw efficiency (a) and absolute increase in efficiency (b) for the bimolecular system (PdOEP+DPA).  The raw efficiency (c) and absolute increase in efficiency (d) for the lanthanide-based system (IR806-sensitized oleylamine-coated $\beta$-NaYF$_4$:Yb,Er nanoparticles). }
\end{figure}

Figure 4 shows efficiency enhancements as a function of bandgap for different upconverter efficiencies. The lowest upconverter efficiencies used in each plot represent recently measured values (20\% for PdOEP/DPA \cite{Castellano20percent} and .1\% for the lanthanide-based system \cite{IR806andLanth}); the higher upconverter efficiencies are meant to reflect the expected results should ongoing work \cite{AshwinCrescents,MetalEnhancedUC_SuperLattice, MetalEnhancedUC_ParticleSandwich,Polman_UC} lead to more efficient upconversion. Though non-ideal spectral positioning and low efficiencies limit current upconverters, these calculations highlight the promise of this technology. For example, application of the extant 20\% efficient bimolecular upconverter to a 2.6 eV bandgap cell would result in an absolute increase in cell efficiency of .62\%. If bimolecular upconversion efficiency could be increased to between 50\% and 100\%, a 1.53\% to 3.06\% absolute increase in efficiency is expected in the same cell. In contrast, application of the existing lanthanide-based system is not expected to enhance solar cell efficiency due to low upconverter quantum efficiencies. However, if lanthanide upconversion reaches efficiencies of 20\%, 50\%, or 100\%, it is expected that the efficiency of a 1.7 eV bandgap solar cell would be boosted from 28.24\% to 28.77\%, 31.27\%, or 34.38\%, respectively. 


	We have investigated the effects of narrow-band upconversion on the efficiency of a solar cell. Generalizing the Shockley-Queisser limit, we determined that the addition of an ideal upconverter with absorption bandwidths between .1 eV and .5 eV should boost the efficiency of a 1.7 eV bandgap cell from 28.24\% to between 33.52\% and 43.58\%. Further, we found that the optimal absorption peaks for such an upconverter lie at 1.02 eV and 1.47 eV, and similarly mapped out the optimal upconverter parameters for any given cell bandgap. Our case studies indicate that moderate absolute efficiency enhancements of $<$1\% should be expected for existing bimolecular and lanthanide-based upconverters. However, should upconverter quantum yields be boosted beyond their current values, cell efficiency enhancements of several absolute percent are predicted. These calculations highlight the promise of upconversion for photovoltaics, and stress the critical roles of upconverter absorption bandwidth and quantum yield in the push towards technological viability.

\begin{acknowledgments}
We thank Mike McGehee, Alberto Salleo, Tim Burke, Diane Wu, and Sacha Verweij for
insightful discussions. Funding for this research was provided by
the Department of Energy under grant number DE-EE0005331,
Stanford's TomKat Center for Sustainable Energy, Stanford
Global Climate and Energy Project (GCEP), the Robert L. and Audrey S. Hancock Stanford Graduate Fellowship, and the National Science Foundation Graduate Research Fellowship under Grant No. 2012122469. Any opinion, findings, and conclusions or recommendations expressed in this material are those of the authors and do not necessarily reflect the views of the National Science Foundation.
\end{acknowledgments}

\vspace{10 mm}


\begingroup
\renewcommand{\section}[2]{}


\end{document}